\documentclass[journal]{IEEEtran}
\usepackage{cite}
\usepackage{longtable}
\usepackage{amsmath,mathtools,amsfonts,amssymb,amsbsy,bm,paralist,theorem,ifthen,color}
\usepackage{epsfig}
\usepackage{epstopdf}
\usepackage{booktabs}
\newtheorem{Proposition}{Proposition}
\theoremstyle{remark}
\newtheorem{remark}{Remark}

\ifCLASSINFOpdf
\else
\fi

\begin{document}
%
\title{An ADMM-Based Approach to Robust Array Pattern Synthesis}
%
%
%

\author{Jintai~Yang,~
        Jingran~Lin,~
        Qingjiang~Shi,~
        and~Qiang~Li
\thanks{J. Yang, J. Lin, and Q. Li are all with the School of Information and Communication Engineering, University of Electronic Science and Technology of China, Chengdu 611731, China (e-mail: j.t.yang@std.uestc.edu.cn; jingranlin@uestc.edu.cn; lq@uestc.edu.cn). Q. Shi is with the School of Software Engineering, Tongji University, Shanghai 200092, China (e-mail: qing.j.shi@gmail.com). J. Yang and Q. Shi are also with the Shenzhen Research Institute of Big Data, Shenzhen 518172, China.}}

%
%




\maketitle

\begin{abstract}
  In most existing robust array beam pattern synthesis studies, the bounded-sphere model is used to describe the steering vector (SV) uncertainties. In this letter, instead of bounding the norm of SV perturbations as a whole, we explore the amplitude and phase perturbations of each SV element separately, thereby obtaining a tighter SV uncertainty model. Based on this model, we formulate the robust array pattern synthesis problem from the perspective of the min-max optimization, which aims to minimize the maximum side lobe response, while preserving the main lobe response. However, this problem is difficult due to the infinitely many non-convex constraints. As a remedy, we employ the worst-case criterion and recast the problem as a convex second-order cone program (SOCP). To solve the SOCP, we further develop an alternating direction method of multipliers (ADMM)-based algorithm, which is computationally efficient with each step being computed in closed form. Numerical simulations demonstrate the efficacy and efficiency of the proposed algorithm.
\end{abstract}


%
\IEEEpeerreviewmaketitle

\section{Introduction}
The array pattern synthesis technique, which steers the array main lobe direction while suppressing the side lobes, has been widely utilized in many electronic systems \cite{Capon1969J,Boyd1997J,Gershman2010J}. However, it may suffer from severe performance degradation in practice due to various steering vector (SV) imperfections. To alliterate this, the robust pattern synthesis problem has been extensively investigated for decades \cite{Vorobyov2013J}.


Many early robust pattern synthesis studies adopted the simple diagonal loading technique, where a $\ell_2$-regularization term of array beamformer was introduced to provide robustness \cite{Cox1987J,Carlson1988J}. However, without utilizing the SV uncertainty information, its performance is usually unsatisfactory. In order to improve the robust performance, several studies took the SV uncertainty distributions into consideration in the robust pattern synthesis design. For instance, \cite{Vorobyov2003J,Lorenz2005J,Li2003J,Lin2007J,Jiang2018J} employed the bounded-sphere model to describe the SV uncertainties (namely, the $\ell_2$-norm of the SV perturbations is bounded by some given constant), and then formulated the robust pattern synthesis problem based on the minimum variance criterion. Adopting the bounded-sphere SV uncertainty model, the authors of \cite{Zhuang2017J} further addressed a pattern synthesis design with robustness against the correlated SV perturbations. In \cite{Khabbazibasmenj2013J}, the robust pattern synthesis scheme for general-rank signal subspace was developed based on the spherical uncertainty model, where the norm of the perturbation matrix (instead of the perturbation vector) is bounded. In addition to bounding the norm of SV perturbations as a whole, \cite{Yue2016J} modeled the random SV phase error as a Gaussian vector, and considered probabilistic constraints; \cite{Mutapcic2007J} and \cite{Yan2008J} further bounded the element-wise uncertainties in array beamformer and SV, respectively, which finally led to a $\ell_1$-regularization penalty term in the objective of the pattern synthesis optimization problem. The performance difference among distinct SV uncertainty models motivates people to keep seeking a more accurate and practical SV uncertainty description. In addition, although many robust pattern synthesis problems have been shown to be solvable, e.g., by recasting them in the form of second-order cone program (SOCP) \cite{Jiang2018J,Vorobyov2003J,Lorenz2005J,Yan2008J,Liu2003J}, there is still an urgent demand for developing computationally efficient algorithms to make these robust designs applicable in practice.

In this paper, we consider the robust pattern synthesis design based on a new SV uncertainty model, which explores the amplitude and phase perturbations of each SV element separately, instead of bounding the $\ell_2$-norm of the SV perturbations as a whole. This model is more intuitive and physically-meaningful than the element-wise uncertainty model considered in \cite{Yan2008J}. Meanwhile, under the same SV perturbation settings, it gives a tighter SV uncertainty set than the commonly used bounded-sphere model, thereby possessing the potential of achieving better robust performance. According to this model, the robust pattern synthesis design is formulated as a min-max problem. By optimizing the array beamformer, we aim to minimize the maximum side lobe response, while preserving the main lobe response. To handle this challenging non-convex problem, we first connect it with the robust problem in \cite{Yan2008J}, and then employ the similar approach to recast it as a convex SOCP. From there, a computationally efficient algorithm is developed based on the alternating direction method of multipliers (ADMM) \cite{Boyd2011J,Lin2016J,Lin2014C,Shen2012J}. The proposed algorithm is particularly suitable for practical implementation by coming up with a simple closed-form solution in each step.

%
%

\section{System Model and Problem Formulation}
Consider an array consisting of $N$ isotropic sensor elements. The presumed array SV associated with angle $\theta$ is denoted as $\hat{\mathbf{a}}_\theta = [\hat{a}_{\theta,1}, \hat{a}_{\theta,2}, \cdots, \hat{a}_{\theta, N}]^T \in \mathbb{C}^{N \times 1}$. Under the far-field assumption, the elements of $\hat{\mathbf{a}}_\theta$ have equal amplitude and only differ in phase. Without loss of generality, we assume $|\hat{a}_{\theta,1}| = |\hat{a}_{\theta,2}| = \cdots = |\hat{a}_{\theta,N}| = 1$.

Let $\mathbf{w} = [w_1, w_2, \cdots, w_N]^T \in \mathbb{C}^{N \times 1}$ be the array beamformer, and the array pattern with angle $\theta$ is given by $|\mathbf{w}^\dag\hat{\mathbf{a}}_\theta|$, where $\mathbf{w}^\dag$ denotes the Hermitian of $\mathbf{w}$. In this paper, we aim to maintain the unit main lobe response in direction $\theta_0$, while suppressing the response in the side lobe region, denoted by $\boldsymbol{\Theta} = \{\theta_1,\theta_2,\cdots,\theta_M \}$, as small as possible. Given the perfect knowledge of SV, such an array pattern can be synthesized by solving the following min-max problem \cite{Lebret1997J}
\begin{subequations}\label{prob:original}
\begin{align}
    \min_{\mathbf{w}}\,&\max_{\theta\in\boldsymbol{\Theta}} \, |\mathbf{w}^\dag\hat{\mathbf{a}}_\theta| \label{obj:orignal} \\
    \qquad {\rm s.t.} ~ &|\mathbf{w}^\dag\hat{\mathbf{a}}_{\theta_0}| \geq 1. \label{cstr:original}
\end{align}
\end{subequations}
Nevertheless, this approach is very sensitive to the mismatches between the presumed and the actual SVs, which are almost inevitable in practice because of element position errors, distorted sensor calibration, and structural scattering, etc. We denote the actual SV with angle $\theta$ as ${\mathbf{a}}_\theta = [{a}_{\theta,1}, {a}_{\theta, 2}, \cdots, {a}_{\theta,N}]^T \in \mathbb{C}^{N \times 1}$, where ${a}_{\theta,n}$ is modeled as a perturbing version of $\hat{a}_{\theta,n}$, with the perturbations appearing in both amplitude and phase. In particular, ${a}_{\theta,n}$ is expressed as
\begin{equation} \label{eq:steering_vec}
  {a}_{\theta,n} = (1 + \Delta u_{\theta,n})\exp(j\Delta \phi_{\theta,n})\cdot \hat{a}_{\theta,n}
\end{equation}
where $\Delta u_{\theta,n}$ and $\Delta\phi_{\theta,n}$ are the random amplitude and phase perturbations of $\hat{a}_{\theta,n}$, respectively. In addition, these two types of perturbations are assumed to be bounded by $ U_{n}$ and $ \Phi_{n}$, respectively, i.e., $|\Delta u_{\theta,n}| \leq  U_{n}, |\Delta\phi_{\theta,n}| \leq  \Phi_{n}$. In practice, the perturbations cannot be too large to exceed the applicable scope of robust pattern synthesis. Hence, we assume $ U_{n} \ll 1$ and $\Phi_{n} \ll \frac{\pi}{2}$. Under this setting, the uncertainty set for ${\mathbf{a}}_{\theta}$ can be defined as
\begin{align}
    \mathcal{C}_{\theta} \triangleq \{ {\mathbf{a}}_{\theta} \mid {a}_{\theta,n} &= (1 + \Delta u_{\theta,n})\exp(j\Delta \phi_{\theta,n}) \cdot \hat{a}_{\theta,n}, \nonumber \\
     &~~~~|\Delta u_{\theta,n}| \leq  U_{n},~|\Delta\phi_{\theta,n}| \leq \Phi_{n},\,\forall\,n\}.
\end{align}

Based on the uncertainty set $\mathcal{C}_\theta$, the robust pattern synthesis problem can be formulated as
\begin{subequations} \label{prob:robust}
\begin{align}
    \min_\mathbf{w}&\max_{\{\theta \in \boldsymbol{\Theta},\, \mathbf{a}_\theta \in \mathcal{C}_\theta\}} |\mathbf{w}^\dag\mathbf{a}_\theta| \label{obj:robust}\\
    {\rm s.t.}\,&\quad~~ |\mathbf{w}^\dag\mathbf{a}_{\theta_0}| \geq 1, ~\forall ~\mathbf{a}_{\theta_0} \in \mathcal{C}_{\theta_0}. \label{cstr:robust}
\end{align}
\end{subequations}
Unfortunately, problem \eqref{prob:robust} is challenging due to the fan-like (non-convex) set $\mathcal{C}_\theta$ and the infinitely many constraints, which motivates us to pursue an efficient approximate solution to it. To this end, we first relax $\mathcal{C}_\theta$ to a convex set $\tilde{\mathcal{C}}_\theta$ as shown in Proposition \ref{proposition:1}.
\begin{Proposition} \label{proposition:1}
Define the SV uncertainty set $\tilde{\mathcal{C}}_\theta$ as
\begin{align}
    \tilde{\mathcal{C}}_\theta = \{{\mathbf{a}}_{\theta} \mid {a}_{\theta,n} = \hat{a}_{\theta,n} + \Delta a_{\theta, n},~& |\Delta a_{\theta,n}| \leq \delta_{n},\, \forall \, n\}\label{eq:Crelax}
\end{align}
with $\delta_{n} = \sqrt{(1+ U_{n})^2 - 2\left(1+ U_{n}\right)\cos\Phi_{n} + 1 }$. Then, any ${\mathbf{a}}_{\theta}$ satisfying \eqref{eq:steering_vec} must belong to $\tilde{\mathcal{C}}_\theta$. Further, ${\mathbf{a}}_{\theta}$ lies on the boundary of $\tilde{\mathcal{C}}_\theta$ if there exists some $n$ such that $\Delta u_{\theta,n} = U_{n}$ and $\Delta\phi_{\theta,n} = \Phi_{n} $ .
\end{Proposition}

\noindent \textbf{Proof.} Notice that \eqref{eq:steering_vec} can be reformulated as
\begin{equation}
    {a}_{\theta,n} = \hat{a}_{\theta,n} + \Delta a_{\theta,n},
\end{equation}
where $\Delta a_{\theta,n} = [(1+ \Delta u_{\theta,n})\exp(j\Delta\phi_{\theta,n}) - 1]\hat{a}_{\theta,n}$. Then, $\Delta a_{\theta,n}$ can be bounded as
\begin{align}
	|\Delta a_{\theta,n}|^2 &=  (1+\Delta u_{\theta,n})^2 - 2(1+\Delta u_{\theta,n})\cos\Delta\phi_{\theta,n} + 1 \nonumber \\
		&\leq  (1+ U_{n})^2 - 2\left(1+ U_{n}\right)\cos\Delta\phi_{\theta,n} + 1 \nonumber\\
		&\leq  (1+ U_{n})^2 - 2\left(1+ U_{n}\right)\cos\Phi_{n} + 1 \nonumber \\
        &\triangleq \delta_{n}^2 \label{eq:delta_a_bound}
\end{align}
where the first ``$\leq$'' is due to the fact that $|\Delta u_{\theta,n}| \leq U_{n}$, and the second ``$\leq$'' comes from the assumption that $|\Delta \phi_{\theta,n}| \leq \Phi_{n} \in [0, \frac{\pi}{2})$. $\hfill{} \Box$


Notice that $\tilde{\mathcal{C}}_\theta$ is actually the element-wise SV uncertainty set considered in \cite{Yan2008J}. By replacing $\mathcal{C}_\theta$ by $\tilde{\mathcal{C}}_\theta$ and following the steps in \cite{Yan2008J}, we can get a convex restriction of problem \eqref{prob:robust}. Specifically, we first employ the worst-case criterion to avoid handling the infinitely many constraints, and then obtain
\begin{subequations} \label{prob:worst_case_1}
\begin{align}
    \min_{\mathbf{w}}~ &~\max_{\theta\in\boldsymbol{\Theta}}\, \max_{\mathbf{a}_\theta \in \tilde{\mathcal{C}}_\theta} \, |\mathbf{w}^\dag\mathbf{a}_\theta| \label{obj:worst_case_1}\\
    {\rm s.t.}~\,&\min_{\mathbf{a}_{\theta_0} \in \tilde{\mathcal{C}}_{\theta_0}}\, |\mathbf{w}^\dag\mathbf{a}_{\theta_0}| \geq 1. \label{cstr:worst_case_1}
\end{align}
\end{subequations}
We next apply the method in \cite{Vorobyov2003J}, and reformulate problem \eqref{prob:worst_case_1} as
\begin{subequations}  \label{prob:worst_case_2}
\begin{align}
    \min_{\mathbf{w}}~&\max_{\theta\in\boldsymbol{\Theta}}~ |\mathbf{w}^\dag\hat{\mathbf{a}}_\theta| + \sum\limits_{n = 1}^N \delta_{n}|w_n|    \label{obj:worst_case_2} \\
    {\rm s.t.}~\,& |\mathbf{w}^\dag\hat{\mathbf{a}}_{\theta_0}| - \sum\limits_{n = 1}^N\delta_{n}|w_n| \geq 1. \label{cstr:worst_case_2}
\end{align}
\end{subequations}
Notice that the min-max problem \eqref{prob:worst_case_2} is still non-convex due to the constraint \eqref{cstr:worst_case_2}. Since rotating the phase of $\mathbf{w}$ does not change the optimality, problem \eqref{prob:worst_case_2} can be equivalently recast as a SOCP \cite{Vorobyov2003J}
\begin{subequations} \label{prob:worst_case_3}
\begin{align}
    \min_{\{\mathbf{w},t\}}~& t + \sum\limits_{n=1}^N\delta_n|w_n| \label{obj:worst_case_3} \\
    {\rm s.t.} ~~ & |\mathbf{w}^\dag\hat{\mathbf{a}}_{\theta_m}| \leq t,~ \forall\, m = 1, 2, \cdots, M, \label{cstr:worst_case_3b}\\
    & \mathbf{w}^\dag\hat{\mathbf{a}}_{\theta_0} \geq \sum\limits_{n=1}^N\delta_n|w_n| + 1,~\Im\{\mathbf{w}^\dag\hat{\mathbf{a}}_{\theta_0}\} = 0, \label{cstr:worst_case_3c}
\end{align}
\end{subequations}
where $t$ is an auxiliary variable, and $\Im\{\cdot\}$ denotes the imaginary part of a complex number.
%

\begin{remark}
    Although \eqref{prob:worst_case_3} is the same as the $\ell_1$-regularization problem in \cite{Yan2008J}, our formulation is derived under a different SV uncertainty model. Originally, our robust problem \eqref{prob:robust} is formulated based on $\mathcal{C}_\theta$. Compared with $\tilde{\mathcal{C}}_\theta$, $\mathcal{C}_\theta$ gives a more intuitive form of the SV element uncertainty which essentially consists of the perturbations in amplitude and phase. These two types of perturbations are usually measurable in practice. Problem \eqref{prob:worst_case_3} is a convex restriction of problem \eqref{prob:robust}. Proposition~\ref{proposition:1} establishes a connection between them by revealing the relation between the total SV element uncertainty, i.e., $\delta_{n}$, and the individual amplitude and phase uncertainties, i.e., $U_{n}$ and $\Phi_{n}$.
\end{remark}

\begin{figure}[tbp]
  \centering
  \includegraphics[width = 5cm]{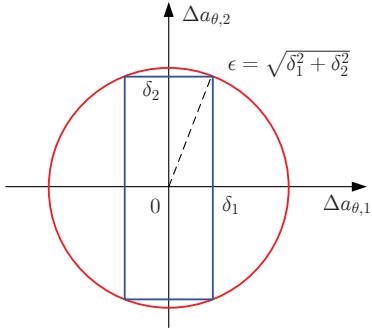}
  \caption{$\tilde{\mathcal{C}}_\theta$ gives a tighter SV uncertainty set than the bounded-sphere model. }
  \label{fig:l1vsl2para}
\end{figure}

\begin{remark}
    As shown in Fig. \ref{fig:l1vsl2para}, $\tilde{\mathcal{C}}_\theta$ gives a tighter SV uncertainty set than the well-known bounded-sphere model, i.e., $\mathcal{B}_\theta = \{{\mathbf{a}}_{\theta} \mid \mathbf{a}_{\theta} = \hat{\mathbf{a}}_{\theta} + \Delta \mathbf{a}_{\theta},~ \|\Delta \mathbf{a}_{\theta}\|_2 \leq \epsilon\}$. Therefore, problem \eqref{prob:worst_case_3} yields a lower worst-case side lobe level.
\end{remark}

So far, we have obtained a convex formulation of the robust pattern synthesis problem, by applying the worst-case criterion and the SOCP reformulation. Problem \eqref{prob:worst_case_3} can be solved by the well known CVX tools \cite{Grant2014M}. On the other hand, to improve the performance of pattern synthesis, there exists a tendency of enlarging the array size $N$ \cite{Lu2014J}. 
Further, to guarantee the side lobe suppression performance, the sample number in the side lobe region, i.e., $M$, cannot be too small. As a consequence, \eqref{prob:worst_case_3} turns out to be a high-dimensional problem in practice. In such circumstances, people prefer a computationally efficient algorithm. To this end, we further recast problem \eqref{prob:worst_case_3} so that it fits into the ADMM framework, and develop a low-complexity algorithm with simple closed-form solutions in each step.%



\section{The ADMM-Based Algorithm}

To make problem \eqref{prob:worst_case_3} amendable to the ADMM framework, we reformulate it as follows:
\begin{subequations} \label{prob:admm}
\begin{align}
    \min_{\{\mathbf{w},\mathbf{v}, \mathbf{x}, t\}}~&t+\sum\limits_{n=1}^N\delta_n|v_n| \label{obj:admm}\\
    {\rm s.t.}\quad~ &|x_{\theta_m}| \leq t,~\forall\, m = 1, 2, \cdots, M, \label{cstr:admm_b} \\
    &x_{\theta_0} \geq \sum\limits_{n=1}^N\delta_n|v_n| + 1,~\Im\{x_{\theta_0}\} = 0, \label{cstr:admm_c}\\
    &x_{\theta_m} = \mathbf{w}^\dag\hat{\mathbf{a}}_{\theta_m},~\forall\, m = 0, 1, \cdots, M, \label{cstr:admm_d} \\
    &\mathbf{v} = \mathbf{w}, \label{cstr:admm_d}
\end{align}
\end{subequations}
where $\mathbf{x} = [x_{\theta_0}, x_{\theta_1}, \cdots, x_{\theta_M}]^T \in \mathbb{C}^{(M+1)\times 1}$ and $\mathbf{v} = [v_1,$ $v_2, \cdots,v_N]^T \in \mathbb{C}^{N\times 1}$  are the introduced auxiliary variables. The equivalence between problems \eqref{prob:worst_case_3} and \eqref{prob:admm} can be easily verified.

By employing the method of \emph{augmented Lagrangian minimization}, problem \eqref{prob:admm} can be handled by solving the following problem
\begin{align}
    \max_{\{\boldsymbol{\lambda}, \boldsymbol{\gamma}\}} ~&\min_{\{\mathbf{w},\mathbf{v}, \mathbf{x}, t\}} \mathcal{L}_\rho(\mathbf{w},\mathbf{v}, \mathbf{x}, t, \boldsymbol{\lambda}, \boldsymbol{\gamma}) \label{prob:aug_lagrange} \\
    {\rm s.t.}~~ &~~\, \eqref{cstr:admm_b}~\text{and}~\eqref{cstr:admm_c}, \nonumber
\end{align}
where $\boldsymbol{\lambda} = [\lambda_0, \lambda_1, \cdots, \lambda_M]^T$ and $\boldsymbol{\gamma} = [\gamma_1, \gamma_2, \cdots, \gamma_N]^T$ are the Lagrangian multipliers; $\rho$ is the positive penalty parameter; $\mathcal{L}_\rho(\mathbf{w},\mathbf{v}, \mathbf{x}, t, \boldsymbol{\lambda}, \boldsymbol{\gamma})$ is the augmented Lagrangian function that is defined as
\begin{align}
    &\mathcal{L}_\rho(\mathbf{w},\mathbf{v}, \mathbf{x}, t, \boldsymbol{\lambda}, \boldsymbol{\gamma}) = t + \sum\limits_{n = 1}^N \delta_n|v_n| \nonumber \\
    &\qquad +  \Re\left\{\sum\limits_{m = 0}^M \lambda_{m}^\ast(x_{\theta_m} - \mathbf{w}^\dag\hat{\mathbf{a}}_{\theta_m}) + \boldsymbol{\gamma}^\dag (\mathbf{v} - \mathbf{w})\right\} \nonumber \\
    &\qquad + \frac{\rho}{2}\left\{\sum\limits_{m = 0}^M |x_{\theta_m} - \mathbf{w}^\dag\hat{\mathbf{a}}_{\theta_m}|^2 + \|\mathbf{v} - \mathbf{w}\|_2^2\right\}, \label{func:aug_lagrange}
\end{align}
where $\lambda_m^\ast$ denotes the conjugate of $\lambda_m, m = 0, 1, \cdots, M$, and $\Re\{\cdot\}$ denotes the real part of a complex number.

By dividing the variables $\{\mathbf{w},\mathbf{v}, \mathbf{x}, t\}$ into two blocks of $\mathbf{w}$ and $\{\mathbf{v}, \mathbf{x}, t\}$, problem \eqref{prob:admm} fits into the framework of ADMM, and hence can be solved by updating $\mathbf{w}$, $\{\mathbf{v}, \mathbf{x}, t\}$, and $\{\boldsymbol{\lambda}, \boldsymbol{\gamma}\}$ alternately. Specifically, the ADMM iterates are given as
\begin{subequations} \label{eq:admm}
\begin{align}
    &\mathbf{w}^{i+1} = \mathop{\rm argmin}_\mathbf{w} \mathcal{L}_\rho(\mathbf{w},\mathbf{v}^i, \mathbf{x}^i, t^i, \boldsymbol{\lambda}^i, \boldsymbol{\gamma}^i) \label{prob:admm_w}\\
    &\{\mathbf{v}^{i+1}, \mathbf{x}^{i+1}, t^{i+1}\} = \mathop{\rm argmin}_{\{\mathbf{v}, \mathbf{x}, t\}}\mathcal{L}_\rho(\mathbf{w}^{i+1},\mathbf{v}, \mathbf{x}, t, \boldsymbol{\lambda}^i, \boldsymbol{\gamma}^i) \label{prob:admm_vxt}\\
    &\qquad\qquad\qquad\qquad\qquad  {\rm s.t.} ~~\eqref{cstr:admm_b}~\text{and}~\eqref{cstr:admm_c}, \nonumber \\
    &\!\begin{cases}
        \lambda_m^{i+1} = \lambda_m^i + \rho[x_{\theta_m}^{i+1} - (\mathbf{w}^{i+1})^\dag\hat{\mathbf{a}}_{\theta_m}],~ \forall\, m\\
        {\gamma}_n^{i+1} = {\gamma}_n^i + \rho({v}_n^{i+1} - {w}_n^{i+1}),~\forall\, n,
    \end{cases}
\end{align}
\end{subequations}
where $i$ is the iteration index.

With the help of ADMM, problem \eqref{prob:admm} is separated into two subproblems, i.e., \eqref{prob:admm_w} and \eqref{prob:admm_vxt}. These two subproblems are very simple and can be solved in closed form, thus generating a computationally efficient algorithm. In the rest of this section, we will elaborate more on the step-by-step computation of the ADMM subproblems.

\subsection{Solution to Subproblem \eqref{prob:admm_w}} \label{subsec:update_w}
Subproblem \eqref{prob:admm_w} is an unconstrained quadratic program (QP) with respect to $\mathbf{w}$, and can be optimally solved in closed form. To simplify the notation, we define
\begin{align}
    \mathbf{A} &\triangleq \sum\limits_{m=0}^M \rho\hat{\mathbf{a}}_{\theta_m}\hat{\mathbf{a}}_{\theta_m}^\dag + \rho\mathbf{I}, \label{eq:matrix_A}\\
    \mathbf{b} &\triangleq \sum\limits_{m=0}^M(\lambda_m + \rho x_{\theta_m}^\ast)\hat{\mathbf{a}}_{\theta_m} + \boldsymbol{\gamma} + \rho\mathbf{v}, \label{eq:vector_b}
\end{align}
and then the subproblem \eqref{prob:admm_w} can be expressed as
\begin{equation}
    \min_{\mathbf{w}}~ \frac{1}{2}\mathbf{w}^\dag\mathbf{A}\mathbf{w} - \Re\{\mathbf{w}^\dag\mathbf{b}\}
\end{equation}
Since $\mathbf{A}$ is positive definite, the optimal $\mathbf{w}$ can be calculated uniquely in closed form, i.e.,
\begin{equation}
    \mathbf{w}^\star = \mathbf{A}^{-1}\mathbf{b}. \label{solution:w}
\end{equation}
Notice that since $\{\hat{\mathbf{a}}_{\theta_m}\}_{m = 0}^M$ are given, $\mathbf{A}^{-1}$ can be computed in advance. The complexities of updating $\mathbf{b}$ and computing $\mathbf{A}^{-1}\mathbf{b}$ are $\mathcal{O}(MN)$ and $\mathcal{O}(N^2)$, respectively. Therefore, the complexity of updating $\mathbf{w}$ in each ADMM iteration is only $\mathcal{O}(MN + N^2)$.

\subsection{Solution to Subproblem \eqref{prob:admm_vxt}} \label{subsec:update_vxt}
The subproblem related to $\{\mathbf{v},\mathbf{x},t\}$ can be expressed as
\begin{align}
    \min_{\{\mathbf{v},\mathbf{x},t\}}~&\left\{\begin{aligned}
    t &+ \sum\limits_{n=1}^N \left(\delta_n |v_n| +\frac{\rho}{2}|v_n + \tfrac{1}{\rho}\gamma_n - w_n|^2\right) \\
    & + \frac{\rho}{2}\sum_{m=0}^M |x_{\theta_m} + \tfrac{1}{\rho}\lambda_m - \mathbf{w}^\dag\hat{\mathbf{a}}_{\theta_m}|^2
    \end{aligned}\right\} \label{prob:admm_vxt2}  \\
    {\rm s.t.}\quad &|x_{\theta_m}| \leq t,~\forall\, m = 1, 2, \cdots, M, \nonumber \\
    &x_{\theta_0} \geq \sum\limits_{n=1}^N\delta_n|v_n| + 1,~\Im\{x_{\theta_0}\} = 0. \nonumber
\end{align}
It can be easily observed that problem \eqref{prob:admm_vxt2} is separable among $\{t,\{x_{\theta_m}\}_{m = 1}^M\}$ and $\{x_{\theta_0}, \{{v}_n\}_{n = 1}^N\}$. Therefore, we can update them independently.

\subsubsection{Update $\{t,\{x_{\theta_m}\}_{m = 1}^M\}$} \label{subsubsec:update_vxt_xt}
To simplify the notation, let us denote $c_{\theta_m} = \frac{1}{\rho}\lambda_m - \mathbf{w}^\dag\hat{\mathbf{a}}_{\theta_m}, m = 1, 2, \cdots, M$. The problem related to $\{t,\{x_{\theta_m}\}_{m = 1}^M\}$ is given by
\begin{align}
    \min_{\{t,\{x_{\theta_m}\}_{m = 1}^M\}}~&t +  \frac{\rho}{2}\sum_{m=1}^M |x_{\theta_m} + c_{\theta_m} |^2 \label{prob:admm_vxt2_tx}  \\
    {\rm s.t.}\qquad~ &|x_{\theta_m}| \leq t,~\forall\, m = 1, 2, \cdots, M. \nonumber
\end{align}
Notice that the constraint $|x_{\theta_m}| \leq t$ does not restrict the phase of $x_{\theta_m}$. To minimize the objective of problem \eqref{prob:admm_vxt2_tx}, $x_{\theta_m}$ must take the following form
\begin{equation}
    x_{\theta_m} = \begin{cases}
        -\frac{c_{\theta_m}}{|c_{\theta_m}|}\cdot |x_{\theta_m}|,\qquad c_{\theta_m} \neq 0, \\
        0,\qquad \qquad\qquad~~~~ c_{\theta_m} = 0.
    \end{cases}
\end{equation}
Consequently, problem \eqref{prob:admm_vxt2_tx} can be simplified as \eqref{prob:admm_vxt2_tx_abs}, which tries to find the optimal $t$ and $z_{\theta_m} = \lvert x_{\theta_m} \rvert, m = 1, 2, \cdots, M$,
\begin{subequations}\label{prob:admm_vxt2_tx_abs}
\begin{align}
    \min_{\{t,\{z_{\theta_m}\}_{m = 1}^M\}}~&t +  \frac{\rho}{2}\sum_{m=1}^M (z_{\theta_m} - |c_{\theta_m}|)^2 \label{obj:admm_vxt2_tx_abs} \\
    {\rm s.t.} \qquad\, & 0 \leq z_{\theta_m} \leq t, ~\forall\, m = 1, 2, \cdots, M. \label{cstr:admm_vxt2_tx_abs_b}
\end{align}
\end{subequations}

Due to the instinct structure, problem \eqref{prob:admm_vxt2_tx_abs} can be optimally solved in closed form \cite{Shi2017J}. First, when $t$ is given in problem \eqref{prob:admm_vxt2_tx_abs}, the optimal solution to $z_{\theta_m}$ can be easily obtained as
\begin{equation}
    z_{\theta_m} = \texttt{Proj}_{[0,t]}\{|c_{\theta_m}|\}, ~\forall\, m = 1, 2, \cdots, M,
\end{equation}
where $\texttt{Proj}_{[0,t]}\{\cdot\}$ denotes the projection onto the range of $[0,t]$. That is, when $|c_{\theta_m}| < t$, $z_{\theta_m} = |c_{\theta_m}|$; otherwise, $z_{\theta_m} = t$. Without loss of generality, we assume $|c_{\theta_1}| \leq |c_{\theta_2}| \leq \cdots \leq |c_{\theta_M}|$. We further assume that there exists some $K$ such that the optimal $t^\star \in (|c_{\theta_{K - 1}}|, |c_{\theta_K}|]$, i.e., $z_{\theta_m} = |c_{\theta_m}|$ for $\theta_1 \leq \theta_m < \theta_K$, and $z_{\theta_m} = t^\star$ for $\theta_K \leq \theta_m \leq \theta_M$. Then, the optimal $t^\star$ can be achieved by
\begin{align}
    t^\star &= \mathop{{\rm argmin}}_{t\geq0}~ t + \frac{\rho}{2}\sum_{m = K}^M (t - |c_{\theta_m}|)^2 \nonumber \\
            &= \left[\frac{\rho\sum_{m = K}^M |c_{\theta_m}| - 1}{\rho(M-K+1)}\right]^+ \triangleq \Gamma(K).
\end{align}
where $[\cdot]^+ = \max\{0,\cdot\}$. Therefore, solving problem \eqref{prob:admm_vxt2_tx_abs} is equivalent to finding the unique $1 \leq K^\star \leq M$ so that $|c_{\theta_{K^\star}}| \geq \Gamma(K^\star) > |c_{\theta_{K^\star - 1}}|$. Then, the optimal solution to problem \eqref{prob:admm_vxt2_tx} is given by
\begin{subequations} \label{solution:xt}
\begin{align}
    &t^\star = \Gamma(K^\star), \\
    &x_{\theta_m}^\star =  \begin{cases}
        - c_{\theta_m}, \qquad ~ 1 \leq m \leq K^\star-1, \\
        - \frac{c_{\theta_m}}{|c_{\theta_m}|}\cdot t^\star,~ K^\star \leq m \leq M.
        \end{cases}
\end{align}
\end{subequations}

\subsubsection{Update $\{x_{\theta_0}, \{{v}_n\}_{n = 1}^N\}$} \label{subsubsec:update_vxt_v}
Again, to simplify the notation, we denote $c_{\theta_0} = \frac{1}{\rho}\lambda_0 - \mathbf{w}^\dag\hat{\mathbf{a}}_{\theta_0}$, and $d_n = \tfrac{1}{\rho}\gamma_n - w_n,~\forall\, n = 1, 2, \cdots, N$. Then, the problem of $\{x_{\theta_0}, \{{v}_n\}_{n = 1}^N\}$ is given by
\begin{align}
    \min_{\{x_{\theta_0}, \{{v}_n\}_{n = 1}^N\}} &\sum\limits_{n=1}^N \left(\delta_n |v_n| +\frac{\rho}{2}|v_n + d_n|^2\right) + \frac{\rho}{2}|x_{\theta_0} + c_{\theta_0}|^2 \label{prob:admm_vxt2_v} \\
    {\rm s.t.}\quad~~  & x_{\theta_0} \geq \sum\limits_{n=1}^N\delta_n|v_n| + 1,~\Im\{x_{\theta_0}\} = 0. \nonumber
\end{align}

It can be easily observed that the constraint in problem \eqref{prob:admm_vxt2_v} does not restrict the phase of $v_n, n = 1, 2, \cdots, N$. Similar as the analysis in Sec. \ref{subsubsec:update_vxt_xt}, we denote $e_n = |d_n| - \frac{1}{\rho}\delta_n$ and $y_n = |v_n|$, and then recast problem \eqref{prob:admm_vxt2_v} equivalently as
\begin{subequations}\label{prob:admm_vxt2_v_abs}
\begin{align}
    \min_{\{x_{\theta_0},\{y_n\}_{n=1}^N\}} ~&(x_{\theta_0} + \Re\{c_{\theta_0}\})^2 + \sum_{n=1}^N (y_n - e_n)^2 \label{obj:admm_vxt2_v_abs} \\
    {\rm s.t.}\qquad\, & x_{\theta_0} \geq \sum_{n=1}^N\delta_n y_n + 1, \label{cstr:admm_vxt2_v_abs_b} \\
    &y_n \geq 0,~\forall\, n = 1, 2, \cdots, N. \label{cstr:admm_vxt2_v_abs_c}
\end{align}
\end{subequations}

By applying the first-order optimality condition, the optimal $x_{\theta_0}^\star$ and $y_n^\star$ are given as
\begin{subequations}
\begin{align}
    &x_{\theta_0}^\star = -\Re\{c_{\theta_0}\} + \tfrac{\xi}{2}, \\
    &y_n^\star = \left[e_n - \tfrac{\xi\delta_n}{2}\right]^+,~\forall\, n = 1, 2, \cdots, N, \label{solution:yn}
\end{align}
\end{subequations}
where $\xi$ is Lagrangian multiplier associated with \eqref{cstr:admm_vxt2_v_abs_b}, which should be carefully chosen such that the Karush-Kuhn-Tucker (KKT) conditions \cite{Boyd2004M} are satisfied.

Specifically, if
\begin{equation}
    -\Re\{c_{\theta_0}\} \geq \sum_{n = 1}^N\delta_n\left[e_n\right]^+ + 1,
\end{equation}
then $\xi = 0$. Otherwise, we find some $\xi > 0$ such that
\begin{equation}
    \tfrac{\xi}{2}-\Re\{c_{\theta_0}\} = \sum_{n = 1}^N\delta_n^2\!\left[\tfrac{e_n}{\delta_n} - \tfrac{\xi}{2}\right]^+ + 1. \label{eq:KKT}
\end{equation}

Without loss of generality, we assume $\frac{e_1}{\delta_1} \leq \frac{e_2}{\delta_2} \leq \cdots \leq \frac{e_N}{\delta_N}$. Then there exists some $L$ such that $y_n = 0$ for $1 \leq n < L$, and $y_n > 0$ for $L \leq n \leq N$. In this case, \eqref{eq:KKT} turns to
\begin{equation}
    \tfrac{\xi}{2}-\Re\{c_{\theta_0}\} = \sum_{n = L}^N\delta_n^2\left(\tfrac{e_n}{\delta_n} - \tfrac{\xi}{2}\right) + 1,
\end{equation}
and $\xi$ is solved as
\begin{align}
    \xi &= \frac{2\left[\sum_{n=L}^N\delta_n e_n+1 +\Re\{c_{\theta_0}\}\right]}{1 + \sum_{n=L}^N \delta_n^2} \\
    &\triangleq \Omega(L). \nonumber
\end{align}
Therefore, to determine $\xi$, we only need to find some $1 \leq L^\star \leq N$ such that $\frac{e_{L^\star -1}}{\delta_{L^\star-1}} \leq \frac{\Omega(L^\star)}{2} < \frac{e_{L^\star}}{\delta_{L^\star}}$. Then, we have
\begin{subequations}\label{solution:x0v}
\begin{align}
    &x_{\theta_0}^\star =  -\Re\{c_{\theta_0}\} + \tfrac{\Omega(L^\star)}{2}, \label{solution:x0}\\
    &v_n^\star = \begin{cases}
        0,~ 1 \leq n \leq L^\star - 1,\\
        -\frac{d_n}{|d_n|}\cdot(e_n - \tfrac{\Omega(L^\star)\delta_n}{2}),~  L^\star \leq n \leq N.
    \end{cases} \label{solution:v}
\end{align}
\end{subequations}
In the special case of $\frac{\Omega(N)}{2} \geq \frac{e_{N}}{\delta_{N}}$, we directly have $x_{\theta_0}^\star = 1$, and $v_n^\star = 0,~ \forall\, n = 1, 2, \cdots, N$.

\subsection{Summary of the ADMM-Based Algorithm} \label{subsec:algo_summary}
At the end of this section, we summarize the ADMM-based algorithm for the robust pattern synthesis problem. The main steps of the algorithm are listed in Table \ref{table:admm_algorithm}, where the stopping criterion is satisfied as the difference between the iterates of two adjacent iterations falls below some given threshold.
\begin{table}[!htbp]
  \caption{Summary of the ADMM-Based Algorithm}
  \label{table:admm_algorithm}
  \centering
  \begin{tabular}{l l}
  \toprule
  1. & Initialize $\{\hat{\mathbf{a}}_{\theta_m}, x_{\theta_m}\}_{m=0}^M$, $\{\delta_n\}_{n=1}^N$, $\mathbf{v}$, $\boldsymbol{\lambda}$, $\boldsymbol{\gamma}$, $\rho$, and $\mathbf{A}^{-1}$;\\ 
  2. & \texttt{\textbf{repeat}}\\
  3. & \quad $\mathbf{w} \leftarrow \mathbf{A}^{-1}\mathbf{b}$; \\
  4. & \quad $t \leftarrow \Gamma(K)$; \\
     & \quad $x_{\theta_m} \leftarrow  \begin{cases}
        - c_{\theta_m}, \qquad ~1 \leq m \leq K-1, \\
        - \frac{c_{\theta_m}}{|c_{\theta_m}|}\cdot t,~ K \leq m \leq M;
        \end{cases}$\\
  5.   & \quad $x_{\theta_0} \leftarrow  -\Re\{c_{\theta_0}\} + \tfrac{\Omega(L)}{2}$; \\
     & \quad $v_n \leftarrow \begin{cases}
        0,~ 1 \leq n \leq L - 1,\\
        -\frac{d_n}{|d_n|}\cdot(|d_n| - \frac{1}{\rho}\delta_n - \tfrac{\Omega(L)\delta_n}{2}),~ L \leq n \leq N;
    \end{cases}$\\
  6. & \quad $\lambda_m \leftarrow \lambda_m + \rho(x_{\theta_m} - \mathbf{w}^\dag\hat{\mathbf{a}}_{\theta_m}),~ 0 \leq m \leq M$,\\
     & \quad ${\gamma}_n \leftarrow {\gamma}_n + \rho({v}_n - {w}_n),~1 \leq n \leq N$;\\
  7. & \texttt{\textbf{until}} some stopping criterion is satisfied; \\
  \bottomrule
  \end{tabular}
\end{table}

\begin{remark}
    As shown in \cite{Fukushima1992J}, if the problem is feasible, and the subproblems in each ADMM iteration can be uniquely solved, then every accumulation point of the iterates generated by the ADMM algorithm is an optimal solution of the problem. Since problem \eqref{prob:worst_case_3} is always feasible, and the subproblems of $\mathbf{w}$ and $\{\mathbf{v},\mathbf{x},t\}$ are uniquely solved by \eqref{solution:w}, \eqref{solution:xt} and \eqref{solution:x0v}, we claim that the proposed ADMM algorithm solves problem \eqref{prob:worst_case_3} optimally.
\end{remark}

The $\mathbf{w}$-minimization step requires the computational cost of $\mathcal{O}(MN+N^2)$ to complete the multiplication of $\mathbf{A}^{-1}\mathbf{b}$. The complexity of the $\{t, \{x_{\theta_m}\}_{m = 1}^N\}$-minimization step is $\mathcal{O}(MN + M\log_2 M)$, which is mainly coming from computing $c_{\theta_m} = \mathbf{w}^\dag \hat{\mathbf{a}}_{\theta_m}$, $ m = 1, 2, \cdots, M$, and the partition-exchange sorter. Similarly, the computational cost of the $\{x_{\theta_0}, \{v_n\}_{n=1}^N\}$-minimization step is $\mathcal{O}(N\log_2 N)$. The complexity of the Lagrangian multipliers updation is $\mathcal{O}(M + N)$. Therefore, the per-iteration complexity of the proposed algorithm is $\mathcal{O}(N^2 + M\log_2M + MN)$. By contrast, solving problem \eqref{prob:worst_case_3} via the interior-point (IP) method \cite{Aharon2001M} requires a complexity of $\mathcal{O}((M+N)^{1.5}N^2)$.

\section{Numerical Simulations}
In this section, we evaluate the performance of the proposed ADMM-based algorithm on a uniform linear array (ULA) with half-wavelength spacing. The desired main lobe direction $\theta_0$ is $90^{\circ}$ and the side lobe region $\boldsymbol{\Theta}$ is specified as $[0^\circ,89^\circ]\cup[91^\circ,180^\circ]$. The presumed SV for angle $\theta$ is defined as $\hat{\mathbf{a}}_{\theta,n} = \exp(j\pi \cos\theta(n-1)),\ \forall\,n = 1, 2, \cdots, N$. All simulations are run on a computer with i5 CPU. The penalty parameter of the ADMM algorithm is set as $\rho = 1$. 

\begin{figure}[tp]
	\centering
	\includegraphics[width = 8cm]{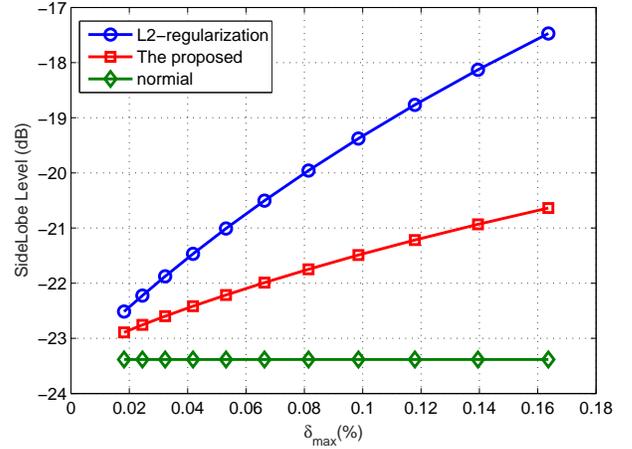}
    \caption{Worst-case side lobe level comparison among the proposed algorithm, the $\ell_2$-regularization approach, and the nominal beamformer approach.}
	\label{fig:l1vsl2result}
\end{figure}

In Fig.\ref{fig:l1vsl2result}, we compare the worst-case side lobe levels of the following three approaches: (1) the proposed approach, (2) the $\ell_2$-regularization approach based on the bounded-sphere SV uncertainty model \cite{Yan2008J}, where the penalty parameter is set as $\epsilon = (\sum_{n=1}^{N} \delta_n^2)^{1/2}$, and (3) the nominal beamformer approach which assumes that the perfect SVs are available. The number of array elements is set as $N = 30$. The side lobe region $\boldsymbol{\Theta}$ is sampled every $1^\circ$, i.e., $M = 180$. 
We set $\Phi_{\max} = 5^\circ$ and sample $U_{\max}$ uniformly from 0.12 to 0.41, which correspond to 1dB and 3dB amplitude perturbations, respectively. Consequently, $\delta_{\max}$ varies from 0.018 to 0.18. We randomly choose $\{U_n, \Phi_n\}_{n = 1}^N$ in the range of $[0, U_{\max}] \times [0, \Phi_{\max}]$. It can be observed from Fig. \ref{fig:l1vsl2result} that 
the proposed approach outperforms the $\ell_2$-regularization approach. This is expected since the SV uncertainty model considered in this paper provides a tighter uncertainty set than the bounded-sphere model, and thus yields a lower worst-case side lobe level.

%
%


Next, we confirm the efficiency advantage of the proposed ADMM-based algorithm. To this end, we solve problem \eqref{prob:worst_case_3} as compared to CVX, which employs the IP method. The CPU running times of the two approaches with different numbers of array elements $N$ and angle samples $M$ are listed in Table \ref{table:time}, where the numerical precision of CVX is $10^{-8}$ and all $\{\delta_n\}_{n=1}^N$ are fixed at 0.15. It can be observed that the proposed algorithm runs much faster than the CVX solver, regardless of the problem dimension.
\begin{table}[h]
	\caption{Comparison of the CPU Running Times}	
    \centering	
	\begin{tabular}{c|c c}
		\toprule
		&\multicolumn{2}{c}{CPU Running Time (Sec.)}\\
		($M$, $N$) & ~ADMM CPU Time~ &~CVX CPU Time~ \\ \hline \hline
        (30, 16) & 0.04  & 0.95 \\
		(60, 30) & 0.06  & 1.55 \\
		(90, 30) & 0.11  & 1.95 \\
		(180, 80) & 0.29 & 4.64 \\
		(360, 200) & 3.29  & 27.54\\
		(720, 500) &  18.65 & 259.77\\
		(1440, 1120) &  92.54 & 2497.87\\
    \bottomrule
	\end{tabular}
	\label{table:time}
\end{table}

\section{Conclusion}
In this paper, we consider the classic robust array pattern synthesis problem, which aims to minimize the maximum side lobe response while preserving the unit main lobe response in the presence of SV imperfections. Departing from most existing studies based on the the bounded-sphere SV uncertainty model, we employ a different uncertainty model by exploring the amplitude and phase perturbations of each SV element separately. One advantage of our strategy lies in that this model provides a tighter SV uncertainty set than the bounded-sphere model. To tackle the infinitely many non-convex constraints within the uncertainty set, we apply the worst-case criterion and recast the min-max problem as a convex SOCP. Then, an ADMM-based algorithm is developed to solve the SOCP problem. The proposed algorithm is computationally efficient since each step can be computed in closed form. Its efficiency and efficacy have been validated by numerical simulations.


%

%
%
%

%

\ifCLASSOPTIONcaptionsoff
  \newpage
\fi


\begin{thebibliography}{10}

\bibitem{Capon1969J}
J. Capon, ``High resolution frequency-wavenumber spectrum analysis,'' {\it Proc. IEEE}, vol. 57, pp. 1408--1418, Aug. 1969.


\bibitem{Boyd1997J}
H. Lebret and S. Boyd, ``Antenna array pattern synthesis via convex optimization,'' {\it IEEE Trans. Signal Process.}, vol. 45, no. 3, pp. 526--532, Mar. 1997.

\bibitem{Gershman2010J}
A.B. Gershman, N.D. Sidiropoulos, S. Shahbazpanahi, M. Bengtsson, and B. Ottersten, ``Convex optimization-based beamforming,'' {\it IEEE Signal
Process. Mag.}, vol. 27, no. 4, pp. 62--75, May 2010.



\bibitem{Vorobyov2013J}
S.A. Vorobyov, ``Principles of minimum variance robust adaptive beamforming design,'' {\it Signal Process.}, vol. 93, no. 12, pp. 3264--3277,
Dec. 2013.



%
%
%
\bibitem{Cox1987J}
H. Cox, R. Zeskind, and M. Owen, ``Robust adaptive beamforming,'' {\it IEEE Trans. Acoustics, Speech, and Signal Processing}, vol. 35, no. 10, pp. 1365--1376, Nov. 1987.

\bibitem{Carlson1988J}
B.D. Carlson, ``Covariance matrix estimation errors and diagonal loading in adaptive arrays,'' {\it IEEE Trans. Aerospace and Electronic systems}, vol. 24, no. 4, pp. 397--401, Aug. 1988.

\bibitem{Vorobyov2003J}
S.A. Vorobyov, A.B. Gershman, and Z.-Q. Luo, ``Robust adaptive beamforming using worst-case performance optimization: A solution to the signal mismatch problem,'' {\it IEEE Trans. Signal Process.}, vol. 51, no. 2, pp. 313--324, Feb. 2003.

\bibitem{Lorenz2005J}
R.G. Lorenz and S. Boyd, ``Robust minimum variance beamforming,'' {\it IEEE Trans. Signal Process.}, vol. 53, no. 5, pp. 1684--1696, May 2005.

\bibitem{Li2003J}
J. Li, P. Stoica, and Z. Wang, ``On robust Capon beamforming and diagonal loading,'' {\it IEEE Trans. Signal Process.}, vol. 51, no. 7, pp. 1702--1714, Jul. 2003.


\bibitem{Lin2007J}
J. Lin, Q. Peng, and H. Shao, ``On diagonal loading for robust adaptive beamforming based on worst case performance optimization,'' {\it ETRI J.}, vol. 28, no. 1, pp. 50--58, Feb. 2007.

\bibitem{Jiang2018J}
X. Jiang, J. Chen, H.C.So, and X.Liu, ``Large-scale robust beamforming via $\ell_\infty$-minimization,'' {\it IEEE Trans. Signal Process.}, vol. 66, no. 14, pp. 3824--3837, Jul. 2018.

\bibitem{Zhuang2017J}
J. Zhuang, B. Shi, X. Zuo, and A. HusseinAli, ``Robust adaptive beamforming with minimum sensitivity to correlated random errors,'' {\it Signal Process.}, vol. 131, pp. 92-98, Feb. 2017.

\bibitem{Khabbazibasmenj2013J}
A. Khabbazibasmenj and S.A. Vorobyov, ``Robust adaptive beamforming for general-rank signal model with positive semi-definite constraint via POTDC,'' {\it IEEE Trans. Signal Process.}, vol. 61, no. 23, pp. 6103--6117, Dec. 2013.

\bibitem{Yue2016J}
M.-C. Yue, S. X. Wu, A.M.-C. So, ``A robust design for MISO physical-layer multicasting over line-of-sight channel,'' {\it IEEE Signal Process. Lett.}, vol. 23, no. 7, pp. 939--943, Jul. 2016.


\bibitem{Mutapcic2007J}
A. Mutapcic, S.-J. Kim, and S. Boyd, ``Beamforming with uncertain weights,'' {\it IEEE Signal Process. Lett.}, vol. 14, no. 5, pp. 348--351, May 2007.

\bibitem{Yan2008J}
S. Yan and J.M. Hovem, ``Array pattern synthesis with robustness against manifold vectors uncertainty,'' {\it IEEE Journal of Oceanic Engineering}, vol. 33, no. 4, pp. 405--413, Oct. 2008.


\bibitem{Liu2003J}
J. Liu, A.B. Gershman, Z.-Q. Luo, and K.M. Wong, ``Adaptive beamforming with sidelobe control: A second-order cone programming approach,'' {\it IEEE Signal Process. Lett.}, vol. 10, no. 11, pp. 331--334, Oct. 2003.

\bibitem{Boyd2011J}
S. Boyd, N. Parikh, E. Chu, B. Peleato, and J. Eckstein, ``Distributed optimization and statistical learning via the alternating direction method of multipliers,'' {\it Foundations and Trends in Machine Learning}, vol. 3, no. 1, pp. 1--122, 2011.

\bibitem{Lin2016J}
J. Lin, Q. Li, C. Jiang, and H. Shao, ``Joint multirelay selection, power allocation, and beamformer design for multiuser decode-and-forward relay networks,'' \textit{IEEE Trans. Veh. Technol.}, vol. 65, no. 7, pp.~5073--5087, Jul. 2016.

\bibitem{Lin2014C}
J. Lin, Y. Li, and Q. Peng, ``Joint power allocation, base station assignment and beamformer design for an uplink SIMO heterogeneous network,'' in {\it Proc. IEEE ICASSP}, Florence, May 2014, pp. 434-438.

\bibitem{Shen2012J}
C. Shen, T.-H. Chang, K.-Y. Wang, Z. Qiu, and C.-Y. Chi, ``Distributed robust multicell coordinated beamforming with imperfect CSI: An ADMM approach,'' \textit{IEEE Trans. Signal Process.}, vol. 60, no. 6, pp.~2988--3003, Jun. 2012.

\bibitem{Lebret1997J}
H. Lebret and S. Boyd, ``Antenna array pattern synthesis via convex optimization,'' {\it IEEE Trans. Signal Process.}, vol. 45, no. 3, pp. 526--532, Mar. 1997.

\bibitem{Grant2014M}
Grant M, Boyd S.: ``The CVX Users' Guide,'' {\it CVX Research Inc.}, Oct. 24, 2014 [on-line]. http://cvxr.com/cvx/.

\bibitem{Lu2014J}
L. Lu, G.Y. Li, A.L. Swindlehurst, A. Ashikhmin, and R. Zhang, ``An overview of massive MIMO: Benefits and challenges,'' {\it IEEE J. Sel. Topics Signal Process.}, vol. 8, no. 5, pp. 742--758, Oct. 2014.



\bibitem{Boyd2004M}
S. Boyd, and L. Vandenberghe, {\it Convex Optimization}, Cambridge, UK: Cambridge University Press, 2004.


\bibitem{Shi2017J}
Q. Shi, M. Hong, X. Fu, and T.-H. Chang, ``Penalty dual decomposition method for nonsmooth nonconvex optimization,'' [Online]. Available: https://arxiv.org/abs/1712.04767

\bibitem{Fukushima1992J}
M. Fukushima, ``Application of the alternating direction method of multipliers to separable convex programming problems,'' {\it Computational Optimization and Applications}, vol. 1, no. 1, pp. 93--111, 1992.

\bibitem{Aharon2001M}
B.-T. Aharon and N. Arkadi, {\it Lectures on Modern Convex Optimization: Analysis, Algorithms, and Engineering Applications}, Society for Industrial and Applied Mathematics, Philadelphia, PA, USA, 2001.

\end{thebibliography}
\end{document}